\documentclass[twocolumn, amsmath,amssymb,aps,groupedaddress,superscriptaddress,prl]{revtex4-1}

\usepackage{graphicx}% Include figure files

\usepackage{amssymb}
\usepackage{amsmath}
\usepackage[usenames,dvipsnames]{color} %for color
\definecolor{LinkColor}{rgb}{0,0,.5}
\usepackage[colorlinks=true,citecolor=LinkColor,urlcolor=LinkColor]{hyperref}

\usepackage{physics}
\renewcommand\phi\varphi

\begin{document}

\title{Emergent perturbation independent decay of the Loschmidt echo in a many-spin system studied through scaled dipolar dynamics}

\author{C. M. S\'anchez}
\affiliation{Facultad de Matem\'atica, Atronom\'{i}a, F\'{i}sica y Computaci\'on, Universidad Nacional de C\'ordoba, C\'ordoba, Argentina.}
\author{A. K. Chattah}
\affiliation{Facultad de Matem\'atica, Atronom\'{i}a, F\'{i}sica y Computaci\'on, Universidad Nacional de C\'ordoba, C\'ordoba, Argentina.}\affiliation{CONICET - IFEG,  C\'ordoba, Argentina.}
\author{K. X. Wei}
\affiliation{Department of Physics,
 Massachusetts Institute of Technology, Cambridge, MA USA.}\affiliation{Research Laboratory of Electronics, Massachusetts Institute of Technology, Cambridge, MA USA}
\author{L.Buljubasich}
\affiliation{Facultad de Matem\'atica, Atronom\'{i}a, F\'{i}sica y Computaci\'on, Universidad Nacional de C\'ordoba, C\'ordoba, Argentina.}\affiliation{CONICET - IFEG,  C\'ordoba, Argentina.}
\author{P. Cappellaro}
\affiliation{Department of Nuclear Science \& Engineering, Massachusetts Institute of Technology, Cambridge, MA USA}\affiliation{Research Laboratory of Electronics, Massachusetts Institute of Technology, Cambridge, MA USA}
\author{H. M. Pastawski}
\affiliation{Facultad de Matem\'atica, Atronom\'{i}a, F\'{i}sica y Computaci\'on, Universidad Nacional de C\'ordoba, C\'ordoba, Argentina.}\affiliation{CONICET - IFEG,  C\'ordoba, Argentina.}

\begin{abstract}
Evaluating the role of perturbations versus the intrinsic coherent dynamics  in driving  to equilibrium is of fundamental interest to understand quantum many-body  thermalization, in the quest to build ever complex quantum devices. 
Here we introduce a  protocol that scales down the coupling strength in a quantum simulator based on a solid-state nuclear spin  system, leading to a longer decay time $T_2$, while keeping perturbations associated to control error constant. We can monitor quantum information scrambling by measuring two powerful metrics, out-of-time-ordered correlators (OTOCs) and Loschmidt Echoes (LEs). While OTOCs reveal quantum information scrambling involving hundreds of spins,  the LE decay quantifies, via the time scale $T_3$, 
how well the scrambled information can be recovered through time reversal. We find that when the interactions dominate the perturbation, the LE decay rate only depends on the interactions themselves, $T_3\propto T_2$, and not on the perturbation. Then, in an unbounded many-spin system,  decoherence can achieve a perturbation-independent regime, with a rate only related to the local second moment of the Hamiltonian.
\end{abstract}

\maketitle

Extensive efforts are  directed at storing and manipulating quantum
information with the goal of building novel quantum devices~\cite{monroe_quantum_2016}. % ranging from superconducting qubits\cite{Neill195} to trapped ions~\cite{Monroe-nature-2017} and atoms~\cite{Lukin-Nature2017}. 
As in experiments the number of qubits   increases~\cite{Monroe-nature-2017,Lukin-Nature2017,Neill18etal}, more
precise control techniques and a deeper understanding of quantum many-body
 dynamics become imperative. In particular, 
assessing the degree of stability of multi-qubit superpositions is crucial not only for technology but also for fundamental physics,  from statistical mechanics
foundations~\cite{MBL-Basko-Altshuler,Choi-Huse-Bloch-Sci2016} to the quantum information  essence of space-time geometry~\cite{PYHP_2015Info-BH,Cowen_entangledNATURE}. In the neighborhood of black holes,
the elementary entangled qubits inherent to a quantum field theory at the
Planck scale~\cite{Maldacena-Shenker-Standford, swingle_unscrambling_2018}  may be affected by a sort of ``quantum butterfly effect''  similar to the Lyapunov instability known to
manifest in classical chaotic systems~\cite{Larkin-Ovchinikov,jalabert_environment-independent_2001,GalitskiPRL17}.

Of particular interest  is to assess how local information
scrambles through a lattice of interacting spin qubits due to unitary
evolution, and how well we can protect or recover the information from scrambling. In addressing these questions,  a privileged position  is still held by  Nuclear Magnetic Resonance (NMR), a quantum  simulator  that allows to tailor Hamiltonian and even invert its sign,  effectively inverting the arrow of time. In addition, solid-state NMR experiments can access large quantum systems, revealing  emergent behavior in the thermodynamic limit~\cite{levstein_attenuation_1998, danieli_quantum_2007, Alvarez_MBLsci2015, wei_exploring_2018, zangara_loschmidt_2017}.

Here we study the dynamics of a 3D nuclear spin lattice under   XXZ effective Hamiltonians with scaled interaction strengths with the goals to (i) create a complex multi-spin superposition through unitary
dynamics, (ii) evaluate the degree of scrambling, understood as the
progressive delocalization of initially localized information, and (iii)  quantify the degree of controllability of these states. 
To characterize the dynamics we implement time
reversal to measure  out-of-time-order correlators (OTOCs)  and Loschmidt Echoes (LE)~\cite{wisniacki_loschmidt_2012} that reveal  how perturbative  terms affect the performance of time reversal. Both OTOCs and LEs are powerful metrics of information scrambling and equilibration in a many-body systems, inspired by the magic echo~\cite{jalabert_environment-independent_2001, sanchez_quantum_2016} and NMR multiple quantum coherence (MQC) experiments~\cite{baum_nmr_1986,
cho_multispin_2005}.

To discriminate between effects due to the intrinsic system dynamics and experimental errors,  we develop  control sequences that engineer a
Floquet transverse dipolar Hamiltonian whose strength can be scaled down by a factor $\pm \delta$. Inverting the Hamiltonian sign  should ideally refocus the  evolution and undo the fast scrambling of local observables. 
%We used them to probe the dynamics of a 3D $^{1}$H nuclear spin system in poly-crystalline adamantane at room temperature in a high magnetic field. 
%The Hamiltonian dynamics induces  fast scrambling of local observables. In addition, by inverting the sign of the Hamiltonian, thus effectively inverting the arrow of time, we can measure OTOCs, until the information is effectively scrambled among over 2$^{100}$ states of the Hilbert space. 
Although our control protocol is imperfect, thus giving rise to a signal decay, it still 
%Our protocol, although naturally imperfect and thus ,
allows us to vary the relative time scales between experimental imperfections and   the dipolar Hamiltonian-induced scrambling. Surprisingly, when trying to fully recover the initial information through a Loschmidt Echo, we find that this recovery has a decay rate whose lower bound remains tied to the strength of the Hamiltonian, i.e., to the dynamics that we should
have been able to refocus. We will discuss how this result hints at the
emergence of an intrinsic irreversibility in the thermodynamic limit
regardless of the weakness of the perturbation terms.

\vspace{-12pt}
\subparagraph{Controlling the dynamics --}
\begin{figure}[t]
\centering
\includegraphics[width=0.9\linewidth]{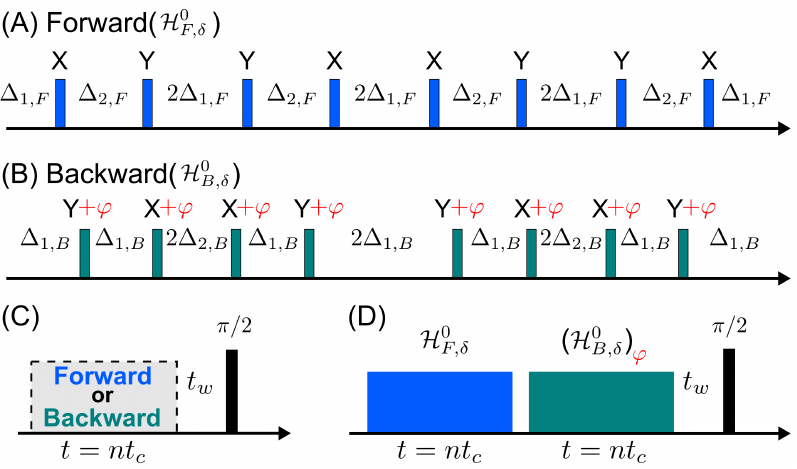}
\caption{\textbf{Experimental Protocol} 
(A) Forward (\textbf{8P$_\delta$-F}) and 
(B) backward (\textbf{8P$_\delta$-B}) pulse sequences  consisting
of 8-$\pi/2$ r.f. pulses with phases as indicated and interpulse
delays  $\Delta^F_{1}\!=\tau(1-\delta)$, $\Delta_{2}^F\!=\tau(1+2\delta)$, $\Delta_{1}^B\!=\tau(1+\delta)$, $\Delta_{2}^B\!=\tau(1-2\delta)$. 16-pulse versions of the sequences can be obtained by repeating the 8-pulse sequence with a $\pi$ shift of all pulses.
(C) Control protocol to probe either forward or
backward dynamics with a scaled dipolar Hamiltonian. The dynamics is probed  measuring the total magnetization $I^z$ by applying a $\pi/2$ pulse after a delay $t_w=500\mu$s that cancels spurious terms. (D) By combining forward and backward evolution  we  observe a Loschmidt echo. Inserting a  variable-phase collective rotation ($\varphi$) allows detecting MQC and extracting OTOCs. 
}
\label{fig:secuencia}
\end{figure}

We consider a system composed by $N$ interacting  spins-$1/2$ (the $^{1}$H nuclear spins in poly-crystalline adamantane at room temperature~\cite{schnell_high-resolution_2001})  in the
presence of a strong magnetic field, $\text{\textbf{B}}_{0}=B_{0}\hat{z}$. In the rotating frame, the Hamiltonian ($\hbar=1$) is 
$\mathcal{H}=-\sum_{i}\omega
_{i}I_{i}^{z}+\mathcal{H}_{\text{d}}^{z}$, where $I_{i}^{\alpha }$ are spin-1/2 operators ($\alpha =x,y,z$).   
The first term is the Zeeman Hamiltonian, with  
 $\omega _{i}$  the difference between the i-th spin's and the
rotating frame frequencies. 
The second term is the secular part of the dipolar Hamiltonian with respect
to the external magnetic field, 
\begin{equation}
\mathcal{H}_{\text{d}}^{z}=\sum_{i<j}d_{ij}(3I_{i}^{z}I_{j}^{z}-\text{%
\textbf{I}}_{i}\cdot \text{\textbf{I}}_{j}),  \label{Dipz}
\end{equation}%
where the dipolar coupling strengths $d_{ij}$ decrease with distance as $r_{ij}^{-3}$.
The system evolves under periodic trains of r.f. pulses that create desired
Floquet Hamiltonians. Specifically, we designed pulse sequences, Fig.~\ref%
{fig:secuencia}(A)-(B), that engineer a scaled dipolar Hamiltonian, with a
scaling factor $\delta$ that can be adjusted by changing the time delays. We refer to the sequences as \textbf{MP$_{\delta }$-A},
where \textbf{A}=\textbf{B},\textbf{F} indicates  backward or forward
respectively, $\delta $ refers to the scaling factor and \textbf{M}$=8$ or $16$ denotes the number of r.f. pulses.
 The \textbf{8P$_\delta$-F,B} (\textbf{16P$_\delta$-F,B}) pulse sequences were designed to have the same number of pulses and cycle time $t_c=12 \tau$ ($24 \tau$), where $\tau$ is a variable time that enforces a safe minimum separation between r.f. pulses. This
ensures that experimental errors are the same across all parameters
employed, for both forward and backward evolution. 
We monitor the evolution with stroboscopic observations at multiples of the cycle time, when the time propagator can be described by an effective time-independent Hamiltonian~\cite{Haeberlen_1976}. This Floquet Hamiltonian can be obtained from a  Magnus expansion $\mathcal{H}_{\delta}=\sum_{i=0}^\infty \mathcal{H}^i_{\delta}$, where $\mathcal{H}^0_{\delta}$ represents the average Hamiltonian and $\mathcal{H}^{i>0}_{\delta}$ are the $i$-th order correction terms. 
The forward (backward) sequence in  Fig.~\ref{fig:secuencia}C(D) yields the average Hamiltonian 
$\mathcal{H}^{0}_{F,\delta}=\delta \,\mathcal{H}^{y}_{d}$ ($\mathcal{H}^{0}_{B,\delta}=-\delta \,\mathcal{H}%
^{y}_{d}$), i.e., a rescaled dipolar Hamiltonian rotated to the $y$ -axis~\cite{Wei18x}.
The scaling factor $\delta$ is in the range $[0,1)$ for the forward sequence, while it can only go up to $1/2$ for the backward sequence that effectively reverses the arrow of time~\footnote{Note that depending on the value of $\tau$ it might not be possible to set $\delta$ to its extreme values, since experimentally we find that there needs to be a minimum time delay between pulses.}.
Both sequences cancel  contributions from the Zeeman Hamiltonian, as well as  other sources of dephasing. 
Furthermore, longer versions  of the sequences (\textbf{16P$_\delta$-F,B}), constructed by simply repeating the cycle with opposite phases,
cancel the first order corrections, $\mathcal{H}_{\delta }^{1}$ (and all other odd-order corrections), as well
as refocusing any heteronuclear dipolar interaction. 

The experiments start with the system initially at equilibrium in a
Boltzmann thermal state, which in the high temperature approximation can be
described by the density operator $\rho (0)=\openone/D+\delta \rho (0)$,
with  $D$ the dimension of the
Hilbert space and $\delta \rho (0)\propto I_z\!=\!\sum_{i}I_{i}^{z}$. As the identity does not evolve nor gives rise to signal, in
the following we will be concerned only with the deviation $\delta \rho $.
The system evolution under a scaled dynamics can be observed by applying the
protocol of Fig~\ref{fig:secuencia}(C):  the system Hamiltonian is quenched to the Floquet Hamiltonian $\pm \delta 
\mathcal{H}_{d}^{y}$ by applying the pulse sequences of Fig.~\ref%
{fig:secuencia}(A-B); then the total magnetization $I^z$ is measured after a short
waiting time $t_{\text{w}}$ by applying a final $\pi /2$-pulse, prior to  signal acquisition (which records the transverse magnetization by
induction). This protocol yields a signal that provides a time-ordered
correlation function: 
\begin{equation}
P_{z}^{\delta }(t)\!=\!\mathrm{Tr\!}\left[ e^{\mathrm{i}t\,\mathcal{H}%
_{\delta }}I^{z}e^{-\mathrm{i}t\,\mathcal{H}_{\delta }}I^{z}\right]
\!=\!\langle I^{z}(t)I^{z}(0)\rangle _{\beta =0}.  \label{Md}
\end{equation}%

\begin{figure}[b]
\centering
\includegraphics[width=0.9\linewidth]{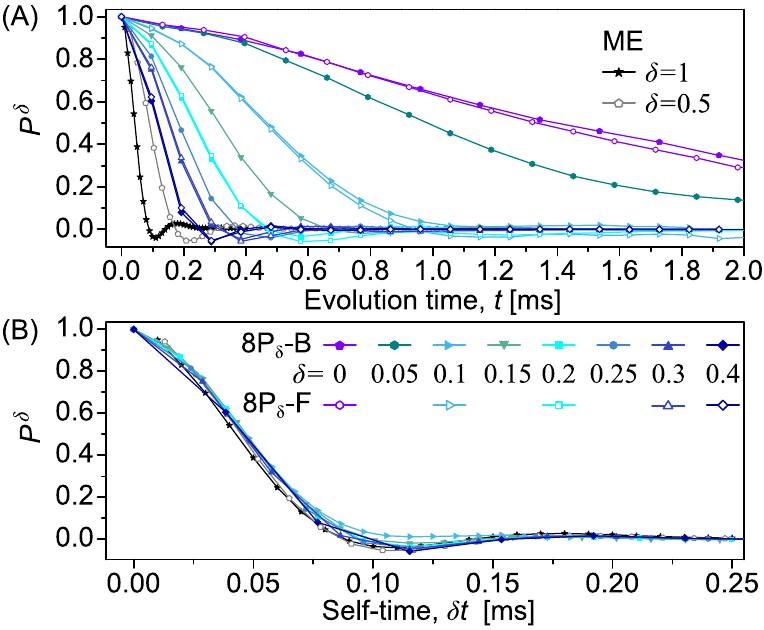}
\caption{Magnetization dynamics for different scaling factors: (A) Comparison of forward and backward
evolutions (and ME) with scaling factors $1$ and $-1/2$, (B) backward
evolutions as a functions of the \textit{self-time}, $\delta t$.
 In the inset: comparison of forward and backward evolutions. }
\label{fig:FID}
\end{figure}

\vspace{-12pt}
\subparagraph{Scaled dynamics --}
To  evaluate the performance of the novel pulses sequences we used this  protocol  to measure the magnetization $P_{z}^{\delta }$ as a function of experimental time $t$  under the \textbf{8P$_{\delta }$-F,B} pulse sequences and scaling in the range $\delta=0-0.4$.
We further consider the case $\delta =1$, obtained by simply allowing free evolution under the natural dipolar Hamiltonian (Eq.~\ref{Dipz}) in between two $\pi /2$-pulses and an additional $\pi$ pulse at half-time, to refocus the Zeeman dynamics.
Also, we include  evolution under a period of high power on-resonance r.f. irradiation of duration $t$ that results in $\delta =-1/2$ (Magic Echo sequence~\cite{Rhim71}). 
The results in Fig.~\ref{fig:FID} clearly show that the dynamics is
modulated by the scaling factor, evolving slower as we decrease $\delta$. We  further check that both forward and backward sequences give rise to
the same magnetization dynamics, a crucial result when one seeks to achieve time reversal to evaluate Loschmidt echo and out-of-time order correlations.  

Ideally we would expect no evolution for $\delta =0$~\footnote{Indeed the sequence becomes a variant of the well-known WHH sequence~\cite
{waugh_approach_1968}, with pulse phases such that also the Zeeman hamiltonian is refocused.}. 
Instead, the observed decay of $P^{\delta =0}(t)$, reveals limitations of the pulse sequence (non-zero higher order terms in the Magnus expansion) and experimental imperfections, due to uncorrected pulse errors. 
We can evaluate the effects of the lowest order errors in the Magnus approximation  by applying symmetrized 16-pulse sequences that exactly cancel out all odd-order terms. 
{While  the \textbf{16P$_{\delta }$-F,B} sequences gives longer coherence times at $\delta=0$, the evolution for $\delta>0$ yields similar results~\cite{SOM}, pointing to the fact that the average Hamiltonian $\mathcal{H}^{0}$ is a good approximation to the effective Hamiltonian, with imperfections dominated by control errors.}
As the longer sequences only allow sparser stroboscopic observations, we kept using the 8-pulse sequences.

To  appreciate the effects of the scaling factor in a more quantitative way, it is useful to plot the magnetizations $P^{\delta }$ as a function of the \textit{self-time}, $\delta t=\delta \times t$, 
Fig.~\ref{fig:FID}(B). Remarkably, all the experimental results collapse to a single curve,   demonstrating the reliability of the \textbf{8P$_{\delta }$-B} pulse sequence, as well as the control used for $\delta =1$ and $\delta =-1/2$. 
For most of the $\delta $ values, the signal is consistent with an evolution dominated by the secular dipolar spin interaction, as highlighted by the fact that the same characteristic oscillation~\cite{abragam_principles_2007} appears around $\delta t\approx100~\mu $s. 
For $\delta =0.1$ and below (not shown), the behavior slightly departs from this common dynamics as we observe a simple decay. We can ascribe this result to the combination of two factors, the longer experimental time elapsed to achieve the same self-time $\delta t$ for smaller $\delta $, and the larger number of pulses required, which introduces more experimental errors. 
The dynamics for larger $\delta $ can instead be accurately fitted by the
well-known function describing the signal decay under dipolar interaction~\cite{abragam_principles_2007},  $P(t)=\textrm{sinc}(wt)e^{-\frac{(ht)^{2}}{2}}$.
The second moment and its corresponding relaxation time $M_{2}=(1/T_{2})^2$ can
be calculated from the fitted parameters, $1/T_{2}=\sqrt{h^{2}+w^{2}/3}$.
We find that $1/T_{2}$ is proportional to $\delta $, that is, the
re-scaled second moment is constant, $1/\delta T_{2}\approx 23.4$kHz. 
This quantitatively confirms the robust behavior of the scaled dynamics 
observed in Fig.~\ref{fig:FID}.

\vspace{-12pt}
\subparagraph{Time reversal --}
\label{mqc} 
The ability to engineer an Hamiltonian with a negative scaling factor  is key to evaluate two quantities, the Loschmidt echo (LE) and the
out-of-time order correlator (OTOC) that can reveal important properties of
the many-spin dynamics. The Loschmidt echo can be observed by combining a
forward and backward sequence (Fig.~\ref{fig:secuencia}.B) for the same time 
$t$ and scaling $\delta $, yielding the signal: 
\begin{equation}
\hspace{-9pt}M^{\delta }\!(t)\!=\!\!\text{Tr}[e^{\mathrm{i}t\mathcal{H}_{F}^{\delta }}e^{%
\mathrm{i}t\mathcal{H}_{B}^{\delta }}I^{z}e^{\textrm{-i}t\mathcal{H}%
_B^\delta}e^{\textrm{-i}t\mathcal{H}_F^\delta}\!I^{z}\!]
\!=\!\langle I_F^{z}(t)I_B^{z}(t)\rangle_0  \label{eq:LE}
\end{equation}%

It is important to remark that if the time inversion were perfect, that is $
\mathcal{H}_{F,\delta }=-\mathcal{H}_{B,\delta }$ to all orders, the
LE intensity would stay constant, $M^{\delta }(t)=1$, thus its decay quantifies intrinsic and experimental imperfections.

\vspace{-12pt}
\subparagraph{OTO correlators and spin counting --} If the fidelity of forward and backward evolutions are good enough, we can
get more insight into the multi-spin correlations created by the dynamics,
widely known in NMR as multiple quantum coherences (MQC)~\cite{baum_multiple-quantum_1985}, by
measuring OTOCs. Note that by inserting a collective spin rotation $\Phi
=e^{-\mathrm{i}\varphi I_{z}}$ between the two 
pulse sequences blocks (see Fig.~\ref{fig:secuencia}D), the signal
is just the OTOC  $S_{\varphi }=\langle {I_{z}(t)\Phi (t)I_{z}(t)\Phi}\rangle
_{\beta =0}$. We can further evaluate the OTO~commutator~\cite
{Wei18x,huang_out--time-ordered_2017, cohn_bang-bang_2018,niknam_2018} 
$\mathcal{C}_{zz}=\langle |[I_{z},I_{z}(t)]|^{2}\rangle _{\beta =0}$,  by repeating the experiment varying $\varphi $ in integer steps, $\varphi
_{n}=2\pi n/Q$, $n=1,...,Q$ and extracting its discrete Fourier transform~
\cite{sanchez_quantum_2016}, 
$S_{q}^{\delta }(t)=\sum_{n=1}^{Q}e^{\mathrm{i}q\varphi }S_{\varphi
_{n}}^{\delta }(t)$.
The MQC intensities $S_{q}$ represent the contribution of all
coherences of order $q$ in the density matrix of the multi-spin state, where
a coherence $q$ with respect to a given basis (here the Zeeman basis $
|{m_z} \rangle$) is any element $\langle r|\rho |s\rangle $ with $s\!-\!r\!=\!q$.
While the sum of MQC intensities yields the LE, their second moment, $
\mathcal{Q}^{2}=\sum_{q}q^{2}S_{q}(t)=\sum_{n}[Q^{2}+Q/\sin (\varphi
_{n}/2)^{2}]S_{\varphi _{n}}(t)$ yields the OTOC~\cite{wei_exploring_2018,Wei18x}, $\mathcal{Q}%
^{2}\propto \mathcal{C}_{zz}$. We remark that this second moment has been
long interpreted in the NMR community as the number of interacting spins
during the forward evolution, $N(t)=\mathcal{Q}$, under the assumption that
due to the strong scrambling driven by  spin-spin couplings all coherence
orders accessible to the interacting spins are fully populated~\cite
{baum_multiple-quantum_1985,baum_nmr_1986}. According to this picture, 
the MQC intensity distribution at a fixed $t$ can be approximated by a
Gaussian, $S(q,N(t))\propto e^{-q^{2}/N^{2}(t)}$, which allows \textit{spin
counting}, that is, evaluating the number of spins that have become
correlated at time $t$~\cite{sanchez_c.m._clustering_2014, sanchez_quantum_2016}. 

We extract the MQC intensities $S_{q}^{\delta }(t)$ by  
implementing the pulse protocol in Fig.~\ref{fig:secuencia}(D) with the 16-pulse sequence for several scaling factors and evolution times. Each MQC distribution was fitted with a  Gaussian to extract its second moment. 
\begin{figure}[t]
\centering
\includegraphics[width=0.9\linewidth]{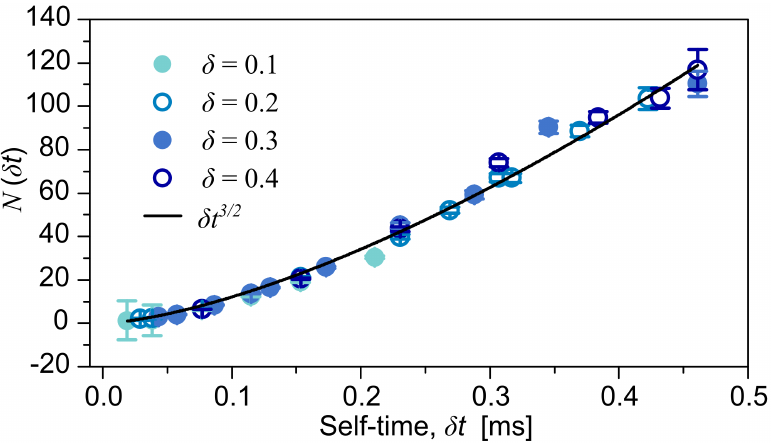}
\caption{Spin counting for the 16- pulse sequence for different scaling
factors as a function of the self-time and the fitting to a power law, 
$N(\delta t) = A \,\delta t^b$, resulting in $b=1.49 \pm 0.04$. }\vspace{-12pt}
\label{sc}
\end{figure}
Figure~\ref{sc} shows how the number of correlated spins $N$ (that is, the square root of the OTOC) grows as a function of the self-time. 
Even when varying the scaling factor $\delta$, the growth of spin correlation always shows the same behavior, and we can fit all the data to a single curve 
$N(\delta t)\sim
\delta t^{3/2}$. This suggests a diffusive spread of
scrambling among $2^{100}$ states of the Hilbert space in our adamantane 3D spin systems, while confirming the good performance of the sequence in scaling the dipolar evolution. 
This fast unbounded growth  contrasts with the  linear growth recently observed in linear chains~\cite{wei_exploring_2018} and weakly coupled molecules~\cite{niknam_2018}, and with the saturation obtained under a  non-interacting  Hamiltonian (the double quantum Hamiltonian~\cite{Pastawski95, Madi97}).

\begin{figure}[t]
\centering
\includegraphics[width=0.9\columnwidth]{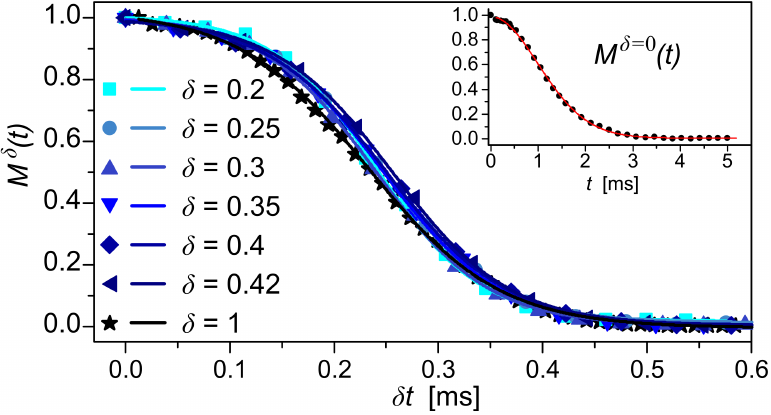}\vspace{-8pt}
\caption{Normalized Loschmidt echoes as a function of the self-time, 
measured following the protocol in Fig.\ref{fig:secuencia}.B, using  the \textbf{8P$_\delta$} pulse sequence, for different scaling factors. The points are data,  lines are Boltzmann function fittings. The inset displays $M^{\delta=0}(t)$. }\vspace{-12pt}
\label{nuevo}
\end{figure}

\vspace{-12pt}
\subparagraph{Loschmidt echoes under scaled dynamics --}
The success of both pulse sequences, \textbf{MP$_\delta$-B,F} in scaling the evolution under the dipolar Hamiltonian is essential to investigate the role of the scaling parameter $\delta$ on the intrinsic irreversibility through the Loschmidt Echo.
Experimental imperfections, such as non ideal r.f. pulses and
 higher order corrections to the average Hamiltonian, introduce a
 perturbative Hamiltonian term $\Sigma$, which is not inverted by our
control and thus produces a decay in the echo intensity as a function of
the evolution time. 
We first consider the LE decay at $\delta\!=\!0$ (inset of Fig.~\ref{nuevo}(B)), which represents the decay only due to these imperfections. 
We fit the $M^{\delta =0}(t)$ decay to the model first proposed in Ref.~\cite
{flambaum-izrailev,zangara_loschmidt_2017} (see~\cite{SOM} for details)
$
f(t)\!=\!\exp\!\left( 2\tfrac{\Gamma ^{2}}{\sigma ^{2}}-2\sqrt{ \tfrac{%
\Gamma ^{4}}{\sigma ^{4}}+\Gamma ^{2}t^{2} }\right)$. 
From this fitting
we define the characteristic time scale of the perturbations $\Sigma$, $T_{\Sigma }$, as the time associated to the half maximum intensity. 

Even for $\delta >0$, the LE should only decay due to imperfections in the
control, while the main dynamics is refocused by the time reversal. Still,
we observe in Fig.~\ref{nuevo} that LE curves, normalized by $M^{\delta =0}(t)$, present decay rates that increase with $\delta$, since they overlap when plotted  as a function of \textit{self-time}. The decays can be best fit to a Boltzmann function (sigmoid), $(1+e^{-\sigma^2(\delta t-t_c)^2})^{-1}$~\cite{SOM}, underlying an initial slow decay.
Strikingly, the overlap of all $M^{\delta }$ as a function of the scaled
time unveils that the rate of intrinsic irreversibility in the many spin
system is directly related to the dominant Hamiltonian dynamics. 
We can define a (scaled) characteristic time for the degradation of the LEs, as a function of self-time, $\widetilde{T}_{3}=T^\delta_3\delta$, given by the time corresponding to  half maximum intensity. 
To confirm our interpretation, we repeated the LE measurements with the
longer \textbf{16P$_{\delta }$} sequence, which cancels the first order
correction arising from finite width pulses ($\propto \delta t_{\pi /2}I_{x}$). 
While 
we obtain a longer $\widetilde{T}_{3}$  time   for the symmetrized sequence ($\sim 20\%$ longer than for \textbf{8P$_{\delta }$}), 
the critical observation that the decay rate is proportional to the scaling $\delta$ still holds.

\begin{figure}[t]
\centering
\includegraphics[width=0.8\linewidth]{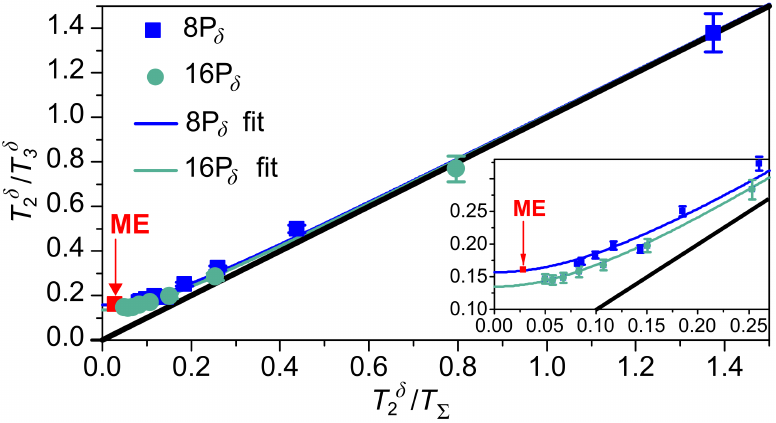}
\caption{Decoherence rate $1/T_{3}^{\delta}$ vs. perturbation rate $1/T_{\Sigma }$,
both multiplied by $T_{2}^{\delta}$, for \textbf{8P$_{\delta }$} (blue)  and \textbf{16P$_{\delta }$} (green). Also shown ME data ($\delta =1$, red). Lines are fittings to the phenomenological function $(T_{2}^{\delta}/T_{3}^{\delta})=\sqrt{R^2+(T_{2}^{\delta}/T_{\Sigma })^2}$.
 The inset displays a zoom in of small value region.}\vspace{-12pt}
\label{T3vsT2}
\end{figure}

Following the analysis in Ref.~\cite{zangara_loschmidt_2017}, we plot the LE decay rate, $1/T_{3}^{\delta }$,  versus the perturbation's characteristic rate $1/T_\Sigma$ for the various pulse sequences we implemented (\textbf{8P$_{\delta }$, 16P$_{\delta }$} and ME). 
Both rates are normalized  by $1/T_{2}^{\delta }$, the second moment of the corresponding evolution  $P^{\delta}(t)$.
When residual interactions (characterized by $T_\Sigma$)  dominate the intrinsic dipolar dynamics (described by $T^\delta_2$) at small $\delta$, the experimental points fall on a line with slope 1 ($T_3^\delta\approx T_\Sigma$). At larger values of $\delta$, instead, (in the limit of vanishing perturbation $T^\delta_2/T_\Sigma\ll1$), the ratio $T_{2}^{\delta }/T_{3}^{\delta }$ surprisingly saturates at a fraction  $R\approx0.15\pm 0.02$ of the reversible time-scale (Fig.~\ref{T3vsT2}). That is, thanks to the good control we achieve in these many body systems,
we can probe a regime where reversible interactions responsible for  information scrambling become dominant over  experimental imperfections and show that, as first proposed in Ref.~\citep{zangara_loschmidt_2017}, it is these interactions responsible for scrambling, which are in principle reversed by the LE, that set the  irreversibility rate revealed by the LE.

\vspace{-12pt}
\subparagraph{Conclusions --}\label{conc}

We  introduced pulse sequences capable to engineer a dipolar
Hamiltonian with a controllable scaling factor and invert its sign. 
We demonstrate the effectiveness of the  pulse sequences in scaling the interactions by showing that the signal following evolution under the scaled dynamics collapses to a single  curve as a function of the \textit{self-time} for a wide range of scaling factors. This is true not only when monitoring local operators via the total magnetization, but also when measuring OTO commutators via  the MQC intensities. Thus the scaled dynamics still drives information scrambling over a large system, and  we observe the number of correlated spins to  grow  with a single diffusive law as a function of  \textit{self-time}.

The good control we demonstrate enables  studying the source  of Loschmidt echo decay, by defining their characteristic decoherence time $T^\delta_{3}$ . 
When the usually dominating dipolar interaction has been canceled out by the control ($\delta=0$), we can extract a direct measure of the perturbation time-scale $T_{\Sigma }$ from the LE decay. 
Remarkably, when the dipolar interaction dominates (large $\delta$), we find $T_{3}^\delta\simeq   T_{2}^\delta/R$, even if the LE should cancel the intrinsic dynamics. This result reinforces the \textit{Central Hypothesis
of Irreversibility} hinted by previous experiments: in a 
many-spin system in the thermodynamic limit, the Loschmidt echo decays with a perturbation-independent rate $1/T_{3}$, that is related to the
local second moment of the unperturbed Hamiltonian, in a similar role to a
Lyapunov exponent, and not to residual experimental imperfections. While this perturbation-independent time scale of the LE decay set a strong limit to preserving quantum information and avoiding thermalization, we note however that the  the decay law  is neither an exponential~\cite{jalabert_environment-independent_2001} nor Gaussian~\cite
{ZurekGaussianLoschmidtEcho}, but closer to a sigmoid~\cite
{rufeil-fiori_effective_2009}. 
This indicates that in many-body systems, the
evolution remains fairly reversible for short times and it is only after
a few times $T_{2}$ that the scrambling dynamics becomes irreversible to all
practical purposes. This initially slow decay opens the possibility to use
error correcting protocols to protect information.

\bibliographystyle{apsrev4-1}
\bibliography{MQCedVF}

\clearpage
\begin{center}
\Large{\textbf{Appendix}}
\end{center}
\appendix

\section*{Experimental Details}
\label{exper}
The spin system consisted of protons in polycrystalline adamantane, a
plastic crystal where the $^1$H spins form a dipole-coupled many-spin
system. In this sample the rapid and isotropic rotation of molecules at room
temperature averages out intramolecular dipole-dipole interactions as well
as chemical shift anisotropy. Further details on polycrystalline adamantane
can be found in Ref.~\cite{schnell_high-resolution_2001}.

All the experiments were performed in a Bruker Avance II spectrometer
operating at 300 MHz Larmor frequency under controlled temperature at $303$~K.

The r.f. power was set to 125 kHz, resulting in   $\pi/2$ pulses
of length $t_{\pi/2}=2 \mu$s. The waiting time $t_w$ before detection was set to $500 \mu $s, in order to allow the decay of unwanted transverse magnetization.
For the backward or forward pulse sequences, the parameter $\tau$ associated
to the inter-pulse spacing, was selected in the range between $6$ to $16 \mu$~s. For a given scaling factor, experiments with different $\tau$ values were conducted. The curves displayed in  Fig.~\ref{Fig3} correspond to
the $\tau$ values producing maximum signal. Evolution times were studied in
the range $0$ to $3$ms. For multiple quantum coherence (MQC) experiments, the phase increments $\varphi$ were varied
in $64$ steps in order to encode up to 32 coherence orders. The scaled
dipolar evolution produces only even coherence orders, although it was
verified the existence of odd coherences originating from  experimental
imperfections, with intensities much lower than the dominant even orders.
Loschmidt echoes and
MQC encoding experiments imply the application of forward and backward
blocks with the same scaling $\delta$, that is in the range of $0$ to $0.42$.
Non-ideality of $\pi/2$ pulses was taken into account.
%%%

\begin{figure}[b!]
\centering
\includegraphics[width=0.9\linewidth]{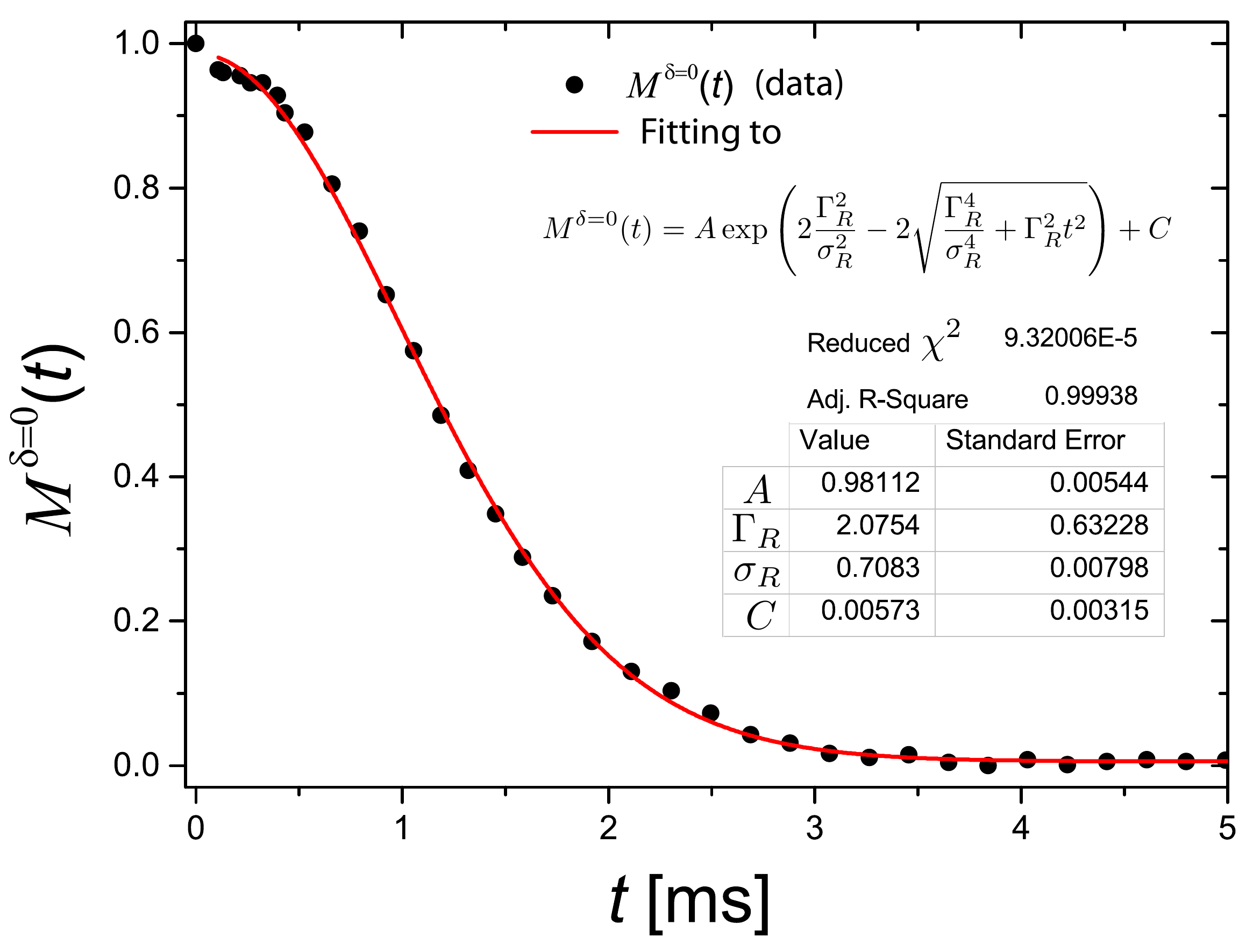}
\caption{Loschmidt echo for $\delta=0$ and the fitting curve.} 
\label{Fig10}
\end{figure}

\section*{Quantifying the residual interaction}

Once the scaling factor vanishes, $\delta=0$, only high order terms in the Magnus expansion remain, including those due to control imperfection. These
residual interactions produce the decay of the Loschmidt echo. 
While a simple Gaussian fit is already a good approximation of the decay, further physical insight, especially about the long-time, exponential behavior, is gained by considering the Flambaum-Izrailev law~\cite{flambaum-izrailev}, which describes both decays:
\begin{equation}
M^{\delta =0}(t)=\exp\left(2\frac{\Gamma_{R}^{2}}{\sigma_{R}^{2}}-2\sqrt{\frac{\Gamma_{R}^{4}%
}{\sigma_{R}^{4}}+\Gamma_{R}^{2}t^{2}}\right)
	\end{equation}
Fitting the experimental data we obtain
\begin{align*}
\Gamma_{R}  & =\left(  2.08\pm0.63\right)  1/ms\\
\sigma_{R}  & =\left(  0.708\pm0.008\right)  1/ms.
\end{align*}
As $\Gamma_R$ is large, at short times the decay is Gaussian, with%
\[
M(t)=\exp[-\sigma_{R}^{2}t^{2}]=\exp[-t^{2}/2T_{\ast}^{2}]
\]
This short time decay has to be compared with the decay of the polarization
due to a forward evolution in a spin environment, using the model proposed by Abragam~\cite{abragam_principles_2007}, which 
quantifies the second moment of the local interactions $M_{2}$
\begin{align*}
P(t)  & =\exp[-a^{2}t^{2}/2]\frac{\sin[bt]}{bt}\\
& \simeq\exp[-M_{2}t^{2}/2]=\exp[-t^{2}/2T_{2}^{2}]
\end{align*}

Experimentally,%
\[
T_{2}=0.03834ms\ll T_{\ast}=\sqrt{2}/0.708=1.998 ms,
\]
which indicates that the scaling pulse sequence plus the Loschmidt time
reversal procedure  has almost frozen  the spin dynamics to the point that the second moment of the residual interactions represents only  2\% of its
original value. Since these residual interactions, being high order, are not
restricted to the original lattice geometry nor to two-body interactions, but involve instead  a huge connectivity in the Hilbert space~\cite{Zangara15}, such a reduction in the
effective second moment is quite remarkable and indicates that the original
interactions have been quite efficiently eliminated. 

On the other hand, for longer times the LE  decays as%
\[
M(t)=\exp[-2\Gamma_{R}t].
\]
This exponential regime can be obtained from a Fermi Golden Rule derivation,
\[
\Gamma_{R}/\hbar=\frac{\pi}{\hbar}\sigma_{R}^{2}\frac
{1}{\sigma_{1R}^{{}}}%
\]
 where $1/\sigma_{1R}=N_{1}$ is the \textbf{Density of Directly
Connected States} of the residual interaction, or correlations function of the
environmental spins, after the scaling and the Loschmidt's time reversal
procedure has been applied. From the experimental fit we obtain
\[\sigma_{1R}=0.759\pm0.232 (ms)^{-1}\]
which then leads to:
\[
\frac{\sigma_{1}}{\sqrt{2}}=0.525(ms)^{-1}
\simeq\frac{1}{T_{\ast}}\equiv\frac{\sigma_{R}}{\sqrt{2}}\ll\frac{1}{T_{2}}
\]

This result  means that  after scaling and time reversal the remaining spin dynamics
can be essentially assigned to a residual network of interacting spins where
the effective interaction itself  have been scaled down by two orders of
magnitude respect to the original interaction. i.e. the scaling is a global
effect. The finding  $\sigma_{1}\simeq\sigma_{R,}$ confirms that Gaussian
decay contains the right physics for all practical purposes. 

\section*{Loschmidt echoes as a function of scaled time, $\delta t$.}
\begin{figure}[h]
\centering
\includegraphics[width=0.99\linewidth]{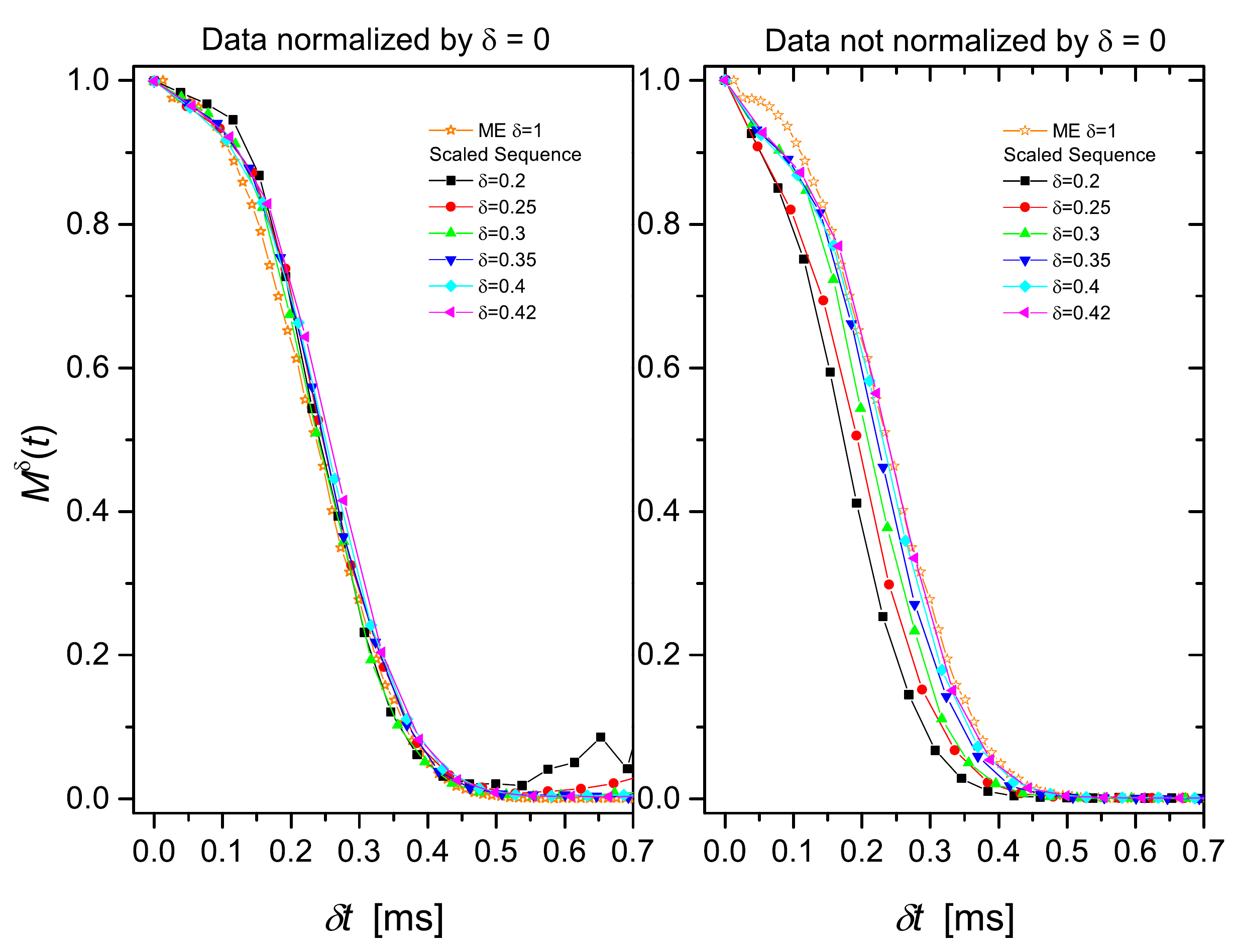}
\caption{Loschmidt echoes as a function of scaled time, $\delta t$. Left side: Normalized $M^{\delta}(t) $ by dividing to $M^{\delta =0}(t) $. Right side: the same data without normalization.} 
\label{Fig1}
\end{figure}

Figure~\ref{Fig1} shows that  only by normalizing to $M^{\delta =0}(t) $  we obtain a single behavior for all scaling factors $\delta $. Indeed, while the left plot shows the overlap of all the curves,  in the right plot we observe instead that the different curves are ordered by the scale factor, decaying faster for smaller $\delta$ factor. This fact can be understood by taking into account that for a given $\delta t$ value, the data corresponding to smaller $\delta$, are obtained with a measurement with a longer experimental time, and therefore more affected by errors, which are removed when normalizing. 

\section*{Comparison of normalized Loschmidt echoes for 8P and 16P sequences}
\begin{figure}[h]
\centering
\includegraphics[width=1.1 \linewidth]{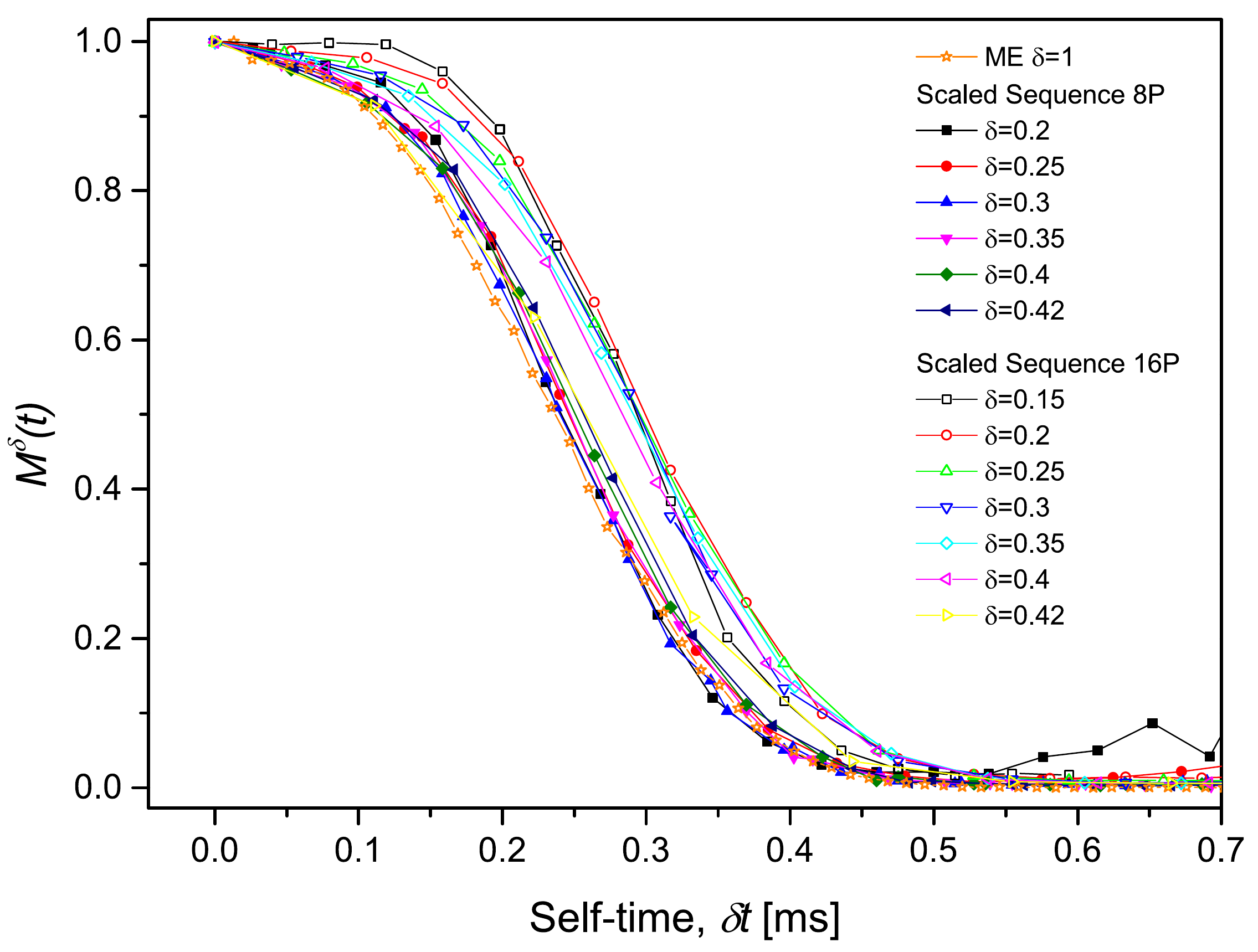}
\caption{Normalized Loschmidt echoes as a function of scaled time, $\delta t$ for 8P and 16P sequence.} 
\label{Fig2}
\end{figure}

Each set of LE curves as a function of $\delta t$  overlap into a single behavior, one for 8P and another  for 16P (see Fig.~\ref{Fig2}). We can observe that the behavior for 16P presents a longer characteristic time for the decay. This fact can be understood taking into account that the 16P sequence cancels out the $\mathcal{H}^{(1)}$ of the AHT. 

\section*{Fitting the Loschmidt Echo curves}
\begin{figure*}[t]
\centering
\includegraphics[width=0.9 \textwidth]{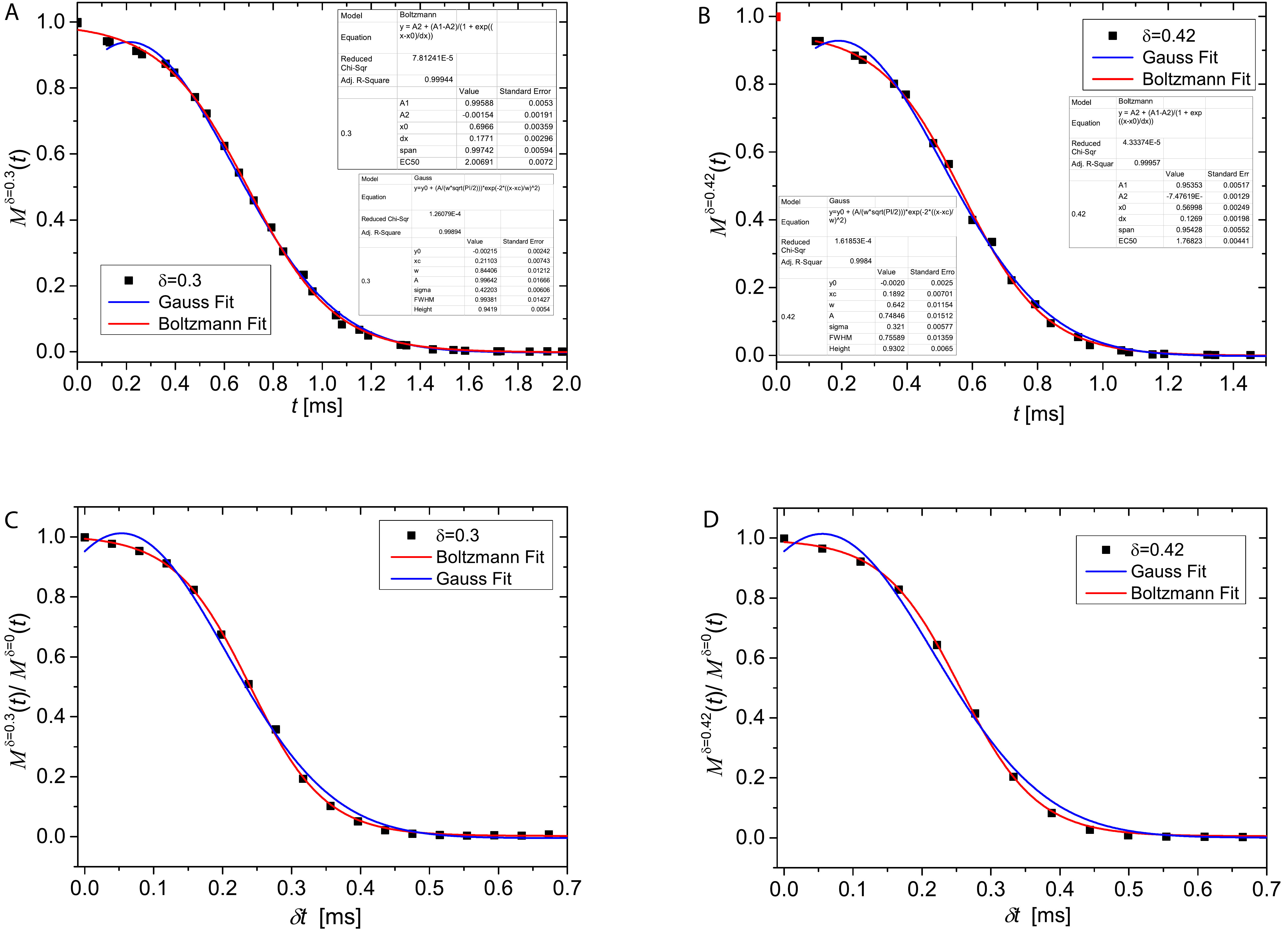}
\caption{Loschmidt Echo without normalization for (A) $\delta =0.3 $ and (B) $\delta=0.42$ and corresponding fitting curves. C-D: Loschmidt Echo with normalization for (C) $\delta =0.3 $ and (D) $\delta=0.42$, and fits.} 
\label{Fig4}
\end{figure*}
We compare models for fitting the LE decay (for $\delta>0$), in particular we consider fitting either to a Gaussian or a Boltzmann function, with equation $ f(t) = A2 + (A1 -A2)/(1+exp((x-x0)/dx))$. 
In figures~\ref{Fig4} A-B, we display, as an example, the LE data without normalization corresponding to $\delta = 0.3$ and $\delta = 0.42$ respectively. 
In the case of non-normalized LEs, fittings with a Gaussian curve or a Boltzmann one are both acceptable, although the Boltzmann one is better. 

Figures \ref{Fig4} C-D display the normalized LE corresponding to the same $\delta = 0.3$ and $\delta = 0.42$ respectively. In these cases we can see that Bolzmann fittings are much better than Gaussian ones.

\section*{Loschmidt echoes as a function of experimental times}

\begin{figure*}[h]
\centering
\includegraphics[width=0.99\linewidth]{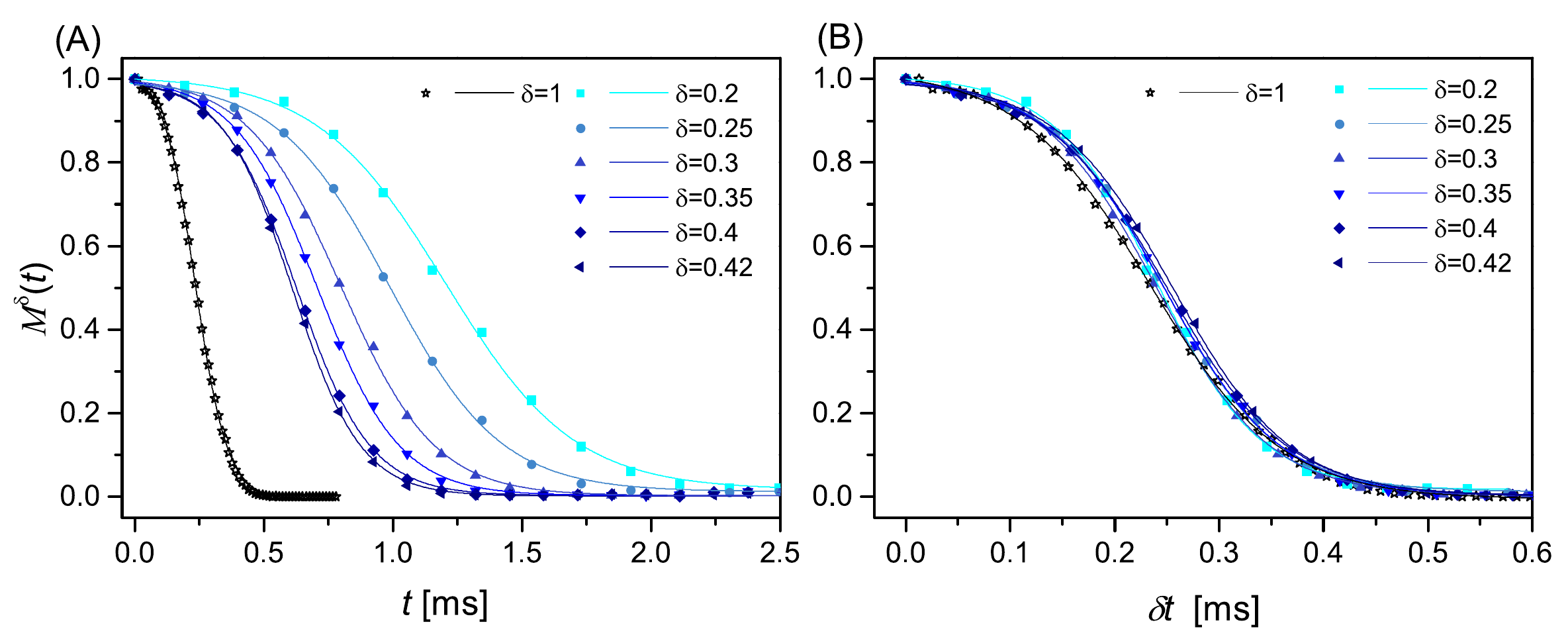}
\caption{ Normalized Loschmidt echoes as a function of (A) experimental time and (B) scaled time, for different $\delta $ performed with the 8P sequence.} 
\label{Fig3}
\end{figure*}

In Fig.\ref{Fig3}(A) we display the normalized LE by dividing to $M^{\delta =0}(t) $, as a function of the experimental time. In this case the curves appear ordered by the scaling factor $\delta$, decaying faster for larger $\delta$ values. In Fig.\ref{Fig3}(B) the same curves plotted vs. the scaled time $\delta t$ overlap in a single behavior.

\section*{Obtaining $T_2$ and $T_3$}

\begin{figure}[h]
\centering
\includegraphics[width=0.99\linewidth]{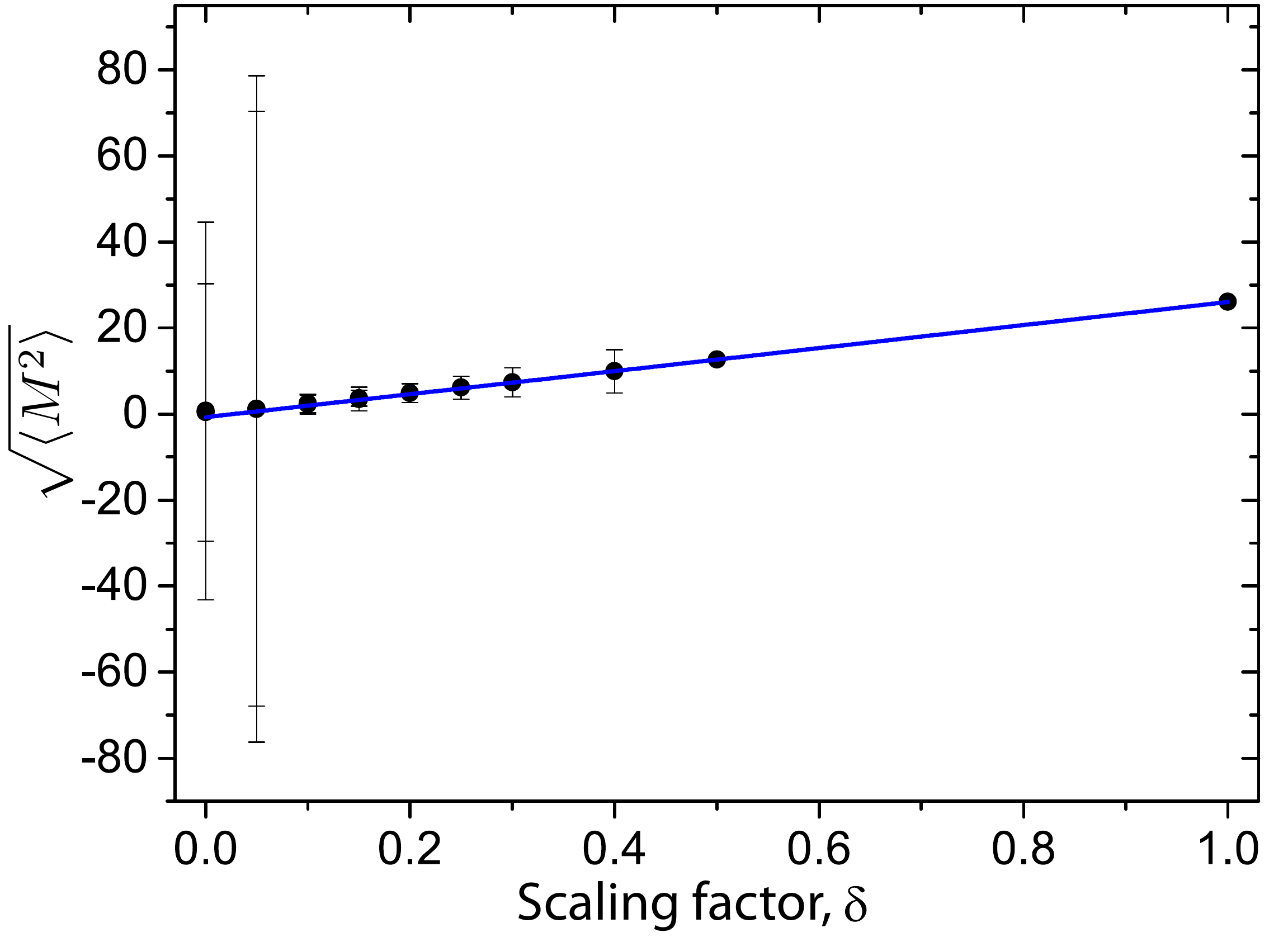}
\caption{Square root of the second moments vs. scaling $\delta$. Dots are data, while the blue line is a linear fit, yielding a slope of 26.77(.06) and intercept -0.71(.03).} 
\label{Fig8}
\end{figure}

The forward or backward dynamics can be fitted by the
well-known function describing the signal decay under dipolar interaction~\cite{abragam_principles_2007},  $P(t)=\textrm{sinc}(wt)e^{-\frac{(ht)^{2}}{2}}$.
The second moment and its corresponding relaxation time $M_{2}=(1/T_{2})^2$ can be calculated from the fitted parameters, $1/T_{2}=\sqrt{h^{2}+w^{2}/3}$. We find that $1/T_{2}$ is proportional to $\delta $, fitting a linear curve, that is shown in Fig.!\ref{Fig8}. 
From this fitting we obtain the $1/T_{2}$ values as a function of $\delta$ used in Fig. 5 of the main paper. 

The $T_3$ values were obtained from the non-normalized LE as a function of experimental time. $T_3$ corresponds to the time associated to the half maximum intensity. Note that these values are independent of the fitting curves mentioned in the previous section. 

The $T_\Sigma$ values were obtained from the LE for $\delta=0$. It corresponds to the time associated to the half maximum intensity. Note that this times are independent of the fitting curve (and it's $T_\Sigma\approx T_\ast$). 

%
%%%
\end{document}